\documentclass[12pt]{article}
\usepackage{a4wide}
\usepackage{graphicx}
\usepackage{amssymb}
\usepackage{axodraw}

\newcommand{\be}{\begin{equation}}
\newcommand{\ee}{\end{equation}}
\newcommand{\ba}{\begin{eqnarray}}
\newcommand{\ea}{\end{eqnarray}}
\newcommand{\lag}{\mathcal{L}}
\newcommand{\mk}{\overline{M}_K^2}
\newcommand{\logm}{\ensuremath{\ell_M}}

\begin{document}

\begin{titlepage}
\begin{flushright}
LU TP 09-14\\
arXiv:0906.0302 [hep-ph]\\
Revised July 2009
\end{flushright}
\vfill
\begin{center}
{\Large\bf \boldmath$K\to\pi\pi$ Decays in $SU(2)$ Chiral Perturbation Theory}
\vfill
{\bf Johan Bijnens and Alejandro Celis}\\[0.3cm]
{Department of Theoretical Physics, Lund University,\\
S\"olvegatan 14A, SE 223-62 Lund, Sweden}
\end{center}
\vfill
\begin{abstract}
We study the decays $K\to\pi\pi$ in one-loop
two-flavour Chiral Perturbation Theory.
We provide arguments why the calculation of the coefficient of the
pionic chiral logarithm
$\logm = M^2\log M^2$
is unique and then perform the calculation. As a check we perform the
reduction of the known three-flavour result. 
Our result can be used to perform the extrapolation to the physical
pion mass of direct lattice QCD calculations of $K\to\pi\pi$
at fixed $m_s$ or $m_K^2$. 
The underlying arguments
are expected to be valid for heavier particles and other processes as well.
\end{abstract}
\vfill
\vfill
\end{titlepage}

\section{Introduction}

Calculating nonleptonic decays precisely from first principles is a
longstanding problem. Progress has been made both on the short distance front
and on the long-distance front. Lattice QCD provides a way to take care
of the latter but is at present limited in the light quark masses that can be
reached. A final extrapolation in the light quark masses is still needed.
For this extrapolation Chiral Perturbation Theory (ChPT)
\cite{Weinberg0,GL0,GL1} is used but in
the nonleptonic sector it has been found that the one-loop corrections for
nonleptonic decays are rather sizable \cite{KMW1,BPP98,BDP}.
The same has also been observed for the quenched and partially
quenched extensions, see e.g. \cite{LC} and references therein.

For static kaon properties like its mass, decay constant and the $B_K$ parameter
an alternative is to use two-flavour ChPT with kaons included.
This was first used for the mass and $\pi K$ scattering in \cite{Roessl},
see also \cite{Ouellette,Frink}, and later extended to the decay constant
and $B_K$ and used for lattice chiral extrapolations \cite{Allton}.
This same method was used for $K_{\ell3}$ at $q^2_{max}$ where the
standard power counting works \cite{Flynn} as well
as for general $q^2$ \cite{Flynn}. In the latter case the standard ChPT
power counting schemes do not work because of the presence of a large momentum
pion. However the authors of \cite{Flynn} argued that also in this case
the coefficient of the chiral logarithm $m_\pi^2\log(m_\pi^2)$ is
calculable.

In this letter we extend the arguments of \cite{Flynn}
to the case of $K\to\pi\pi$ decays and calculate the pionic chiral logarithm
for these decays. We expect that this type of arguments can be applied
to more general processes as well as discussed in Sect.~\ref{powercounting}.
These results are also discussed in the thesis \cite{Celis}.

The expected main use of our result (\ref{resultNLO})  is in extrapolating
lattice QCD results for $K\to\pi\pi$ done at a fixed value of $m_s$
and/or $m_K^2$ in the light quark mass $\hat m$ to the physical pion mass.
This should be possible even when three-flavour ChPT does not work well
since it only requires that two-flavour ChPT is applicable. This is the
main motivation behind this work and the work of \cite{Flynn}.
At present not much data exist directly calculating $K\to\pi\pi$ so
we have not compared our results to lattice data. We hope this will
become feasible in the future. The present status of lattice calculations
relevant for $K\to\pi\pi$ decays is discussed in \cite{LC,Lellouch}.

In Sect.~\ref{twoflavour} we discuss two-flavour ChPT and include
the kaon as a heavy particle \cite{Roessl,Allton} and add the
nonleptonic weak decay
sector to it. Sect.~\ref{powercounting} describes the general argument
why we expect
that also hard pions can be treated  using ChPT and give in particular
the argument for the case of $K\to\pi\pi$. Sect.~\ref{calculation} presents
the results
of the one-loop calculations in two-flavour ChPT while in
Sect.~\ref{threeflavour} we check that the three-flavour result contains
the same logarithms. In Sect.~\ref{conclusions} we summarize our results.

\section{Two-flavour ChPT}
\label{twoflavour}

\subsection{Strong and semileptonic Lagrangian}

Two-flavour ChPT in the meson sector is given in \cite{GL0}.
We use here the exponential notation for the pion field instead.
The notation is the same as in \cite{BCE}.
The lowest order Lagrangian is
\be
\label{pilagrangian}
\lag_{\pi \pi}^{(2)}= \frac{F^2}{4}  
\left( \langle u_{\mu} u^{\mu} \rangle   + \langle \chi_{+}\rangle \right),
\ee
with
\ba
 u_{\mu} &=& i\{  u^{\dag}( \partial_{\mu} - i r_{\mu}   )u
 -u ( \partial_{\mu}   -i l_{\mu}    ) u^{\dag}   \}\,,
\nonumber\\
\chi_{\pm} &=& u^{\dag} \chi u^{\dag} \pm u \chi^{\dag} u\,,
\nonumber\\
u &=& \exp\left(\frac{i}{\sqrt{2}F} \phi \right)\,,
\nonumber\\
 \chi &=& 2B (s+ip) ,\nonumber \\
\phi &=&
 \left( \begin{array}{cc}
\frac{1}{\sqrt{2}}\pi^0  &  \pi^+  \\
 \pi^- & -\frac{1}{\sqrt{2}}\pi^0 \\
 \end{array} \right)\,.
\ea
The field $u$ transforms under a chiral transformation $g_L\times g_R
\in SU(2)_L\times SU(2)_R$ as
\be
u \longrightarrow g_R u h^\dagger = h u g_L^\dagger\,.
\ee
$h$ depends on $u$ and $g_L$, $g_R$ and is the socalled compensator
field. Under this transformation $u_\mu\longrightarrow h u_\mu h^\dagger$.
The notation $\langle X\rangle$ stands for trace over up and down quark
indices and all matrices are $2\times2$ matrices.

We now introduce a kaon field $K$ that is a doublet under isospin 
\be
K = \left(\begin{array}{c}K^+\\ K^0\end{array}\right)\,,
\ee
which transforms under a chiral transformation as
\be
\label{Ktransformation}
K \longrightarrow h K\,.
\ee
We can define a covariant derivative for objects that transform
as (\ref{Ktransformation}) and for those transforming as
$A\longrightarrow h A h^\dagger$ via
\ba
 \nabla_{\mu} A&=& \partial_{\mu}A + [\Gamma_{\mu} ,A  ]\, ,
\nonumber \\
\nabla_{\mu}K &=& \partial_\mu K + \Gamma_{\mu}K\,,
\nonumber\\
\Gamma_{\mu} &=& \frac{1}{2}\{ u^{\dag} (  \partial_{\mu} -i r_{\mu} ) u
 + u(\partial_{\mu} -i l_{\mu}) u^{\dag}   \} \,.
\ea
The fields $s$, $p$ $ r_{\mu} = v_{\mu} + a_{\mu}$ ,
$l_{\mu} =  v_{\mu} - a_{\mu}$ are the standard external scalar, pseudoscalar,
left- and right-handed vector fields introduced by Gasser and Leutwyler.
The mass term for the light quarks is introduced by setting
\be
s = \left(\begin{array}{cc} m_u & \\ & m_d \end{array}\right)\,.
\ee
In this paper we always work in the isospin limit $m_u=m_d=\hat m$.

The effective ChPT Lagrangian contributing to pion-kaon scattering up
to second chiral order is given by \cite{Roessl}
\ba
\label{Klagrangian}
\lag_{\pi K}^{(1)}&=& \nabla_{\mu} K^{\dag} \nabla^{\mu} K
 - \mk K^{\dag} K\,,
\nonumber \\
\lag_{\pi K}^{(2)}&=& A_1 \langle u_{\mu} u^{\mu}  \rangle K^{\dag} K
  + A_2 \langle u^{\mu} u^{\nu}  \rangle \nabla_{\mu} K^{\dag} \nabla_{\nu} K
  + A_3 K^{\dag} \chi_{+} K + A_4 \langle \chi_{+}\rangle K^{\dag}K .
\ea
The chiral order associated with each class of terms corresponds to the
chiral order of the leading tree-contributions and is indicated as an upper
index $(i)$. In (\ref{Klagrangian}) we introduced the notation
$\mk$ for the kaon mass in the limit where $\hat m = 0$.
Similarly we use the $M^2 = 2 B \hat m$ for the lowest order pion mass.

The kaon mass up to order $\hat m$ has no chiral logarithms \cite{Roessl}
and those for the pion mass are well known \cite{GL0}
\ba
M_\pi^2 &=& M^2 \left(1-\frac{1}{2F^2}\overline A(M^2)+2\frac{M^2}{F^2} l_3^r
+\cdots\right)\,,
\nonumber\\
M_K^2 &=& \mk -2 M^2 \left(A_3+2 A_4\right)+\cdots\,. 
\ea
Here we introduced the one-loop function
\be
\overline A(M^2) = -\frac{M^2}{16\pi^2}\log\left(\frac{M^2}{\mu^2}\right)\,.
\ee

The decay constant for the pion is treated in the usual way with \cite{GL0}
\be
\label{Fpi}
F_\pi = F  \left(1+\frac{1}{F^2}\overline A(M^2)+\frac{M^2}{F^2} l_4^r\right)
\,.
\ee
The kaon decay constant needs the introduction of the weak current
\be
\bar s_L \gamma_\mu u_L\,.
\ee
This can be done by introducing a spurion field $t_{L\mu}$ transforming
such that $t_L \longrightarrow g_L t_L^\dagger$ under $SU(2)_L$.
The combination $(t^\dagger_{L\mu})^i \,\bar s_L \gamma^\mu q_{Li}$
with $q_1 = u$ and $q_2 = d$ is then chirally invariant.
The Lagrangian  coupling the kaons is thus given by
\cite{Allton,Flynn}
\be
\label{Kuslagrangian}
\lag_{Kus} = w_1 t^\dagger_{L\mu}u^\dagger\nabla^\mu K 
 + w_2  t^\dagger_{L\mu}u^\dagger u^\mu K + h.c.\,.
\ee
From this one can derive the correction to $F_K$ \cite{Allton}
\be
\label{FK}
F_K = \overline{F}_K\left(1+\frac{3}{8F^2}\overline A(M^2)+\cdots\right)\,.
\ee
$\overline{F}_K$ is the kaon decay constant in the limit $\hat m=0$
and the dots stand for terms of order $\hat m$ but no logarithms.
The terms in (\ref{Kuslagrangian}) are zeroth and first order in the
chiral counting for $F_K$ and $K_{\ell3}$ at $q^2_{max}$.

\subsection{The nonleptonic Lagrangian}

At the quark level the two dominant $\Delta S = -1$ operators are given by
\be
\label{operators}
(\bar s_L\gamma_\mu u_L)(\bar u \gamma^\mu d_L)
\qquad
\mathrm{and}
\qquad
(\bar s_L\gamma_\mu d_L) (\bar u \gamma^\mu u_L)\,.
\ee
We can again makes these terms fully chirally invariant
by adding a spurion $t^{ij}_k$ transforming as
$t^{ij}_k\longrightarrow t^{i'j'}_{k'} =
 t^{ij}_k (g_L)_{k'}^{\phantom{k'}k} (g_L^\dagger)_{i}^{\phantom{i}i^\prime}
(g_L^\dagger)_{j}^{\phantom{j}j^\prime}$.
The term
\be
t^{ij}_k (\bar s \gamma_\mu q_{Li})(\bar q_{L}^k \gamma^\mu q_{Lj})
\ee
is then fully chirally invariant. 
We can actually simplify a little since the operators in (\ref{operators})
transform as a doublet or triplet, $\Delta I=1/2$ or $3/2$, under $SU(2)_L$.
The double combination of the operators can be made invariant
by a single spurion $t_{1/2}$ transforming as $t_{1/2}^i\longrightarrow
t_{1/2}^{i'} = t^{i}_{1/2} (g_L^\dagger)_{i}^{\phantom{i}i^\prime}$.

The actual operators then correspond to the values $t_{1/2}^1 = 0$, $t_{1/2}^2 =1$
for the $\Delta I = 1/2$
and $t^{12}_{1} = t^{21}_1 = - t^{22}_2 = 1$, others zero,
for the $\Delta I = 3/2$ operator.

In constructing possible terms, we can use the identities
$2 u_\mu u^\mu = \langle u_\mu u^\mu\rangle$ and $\langle u_\mu\rangle=0$,
as well as the equations of motion.
When calculating for our case here, i.e. $\chi=\mathrm{diag}(\hat m,\hat m)$,
we have in addition $\langle \chi_-\rangle = 0$ and
$\langle \chi_+\rangle= 2\chi_+$.

We have ordered the terms here by the counting in derivatives
and powers of $\chi$, but how they do
contribute is discussed in Sect.~\ref{powercounting}.

The $\Delta I = 1/2$ terms are using the quantity $\tau_{1/2} = t_{1/2} u^\dagger$
\ba
\label{lag1/2}
\lag_{1/2} &=& iE_1\,\tau_{1/2} K+ E_2 \,\tau_{1/2} u^\mu\nabla_\mu K
+ i E_3 \langle u_\mu u^\mu\rangle \tau_{1/2} K
+ i E_4 \tau_{1/2}\chi_+ K + i E_5 \langle\chi_+\rangle\tau_{1/2} K
\nonumber\\&&
+ E_6 \tau_{1/2}\chi_- K + E_7 \langle\chi_-\rangle\tau_{1/2} K
+ i E_8 \langle u_\mu u_\nu\rangle \tau_{1/2} \nabla^\mu\nabla^\nu K
+ \cdots + h.c.\,.
\ea

By using the equations of motion the first term can be traded for
$\tau_{1/2} \nabla_\mu\nabla^\mu K$. The terms with zero or two derivatives
or one power of $\chi$ are a complete set. We have kept one term
with four derivatives to show that the arguments presented in
Sect.~\ref{powercounting} work for that example. The factors
of $i$ are chosen such that a real coefficient corresponds to
a CP conserving term.

For the $\Delta I = 3/2$ case, we introduce the quantity
$\tau^{ij}_k \equiv t^{i'j'}_{k'}
(u^\dagger)_{i'}^{\phantom{i'}i}(u^\dagger)_{j'}^{\phantom{j'}j}
 u_{k'}^{\phantom{k'}k}$ and get the Lagrangian to second order in derivatives
or first order in $\chi$
\ba
\label{lag3/2}
\lag_{3/2}&=&
iD_1\tau^{ij}_k (u_\mu)_{i}^{\phantom{i}k}(u^\mu K)_j  
+D_2\tau^{ij}_k (u_\mu)_{i}^{\phantom{i}k}(\nabla^\mu K)_j  
+iD_3\tau^{ij}_k (\chi_+)_{i}^{\phantom{i}k}  K_j 
\nonumber\\&& 
+D_4\tau^{ij}_k (\chi_-)_{i}^{\phantom{i}k}  K_j 
+\cdots+h.c.\,. 
\ea
A term like $i\tau^{ij}_k (u_\mu u^\mu)_{i}^{\phantom{i}k}  K_j$
never contributes since $t^{ij}_k$ is such that the trace part of
the first factor does not contribute. This also means that in the isospin
limit the $D_3$ and $D_4$ terms never contribute.
Here we have not included any terms with more derivatives. 

\section{An argument why $K\to\pi\pi$ can be treated}
\label{powercounting}

{\bf 1.)} 
A general reason why we expect that there might be some predictions possible
also for processes with large momentum pions is that chiral logarithms
are caused by small momentum pion propagators. Soft pion
couplings are related directly using the soft pion theorem,
\be
\label{currentalgebra}
\lim_{q\to 0}\langle\pi^k(q)\alpha|O|\beta\rangle
 = -\frac{i}{F_\pi}\langle\alpha|\left[Q_5^k,O\right]|\beta\rangle\,,
\ee
to matrix elements without the soft pion.
The states $\alpha$ and $\beta$ can also contain large momentum
pions. The underlying problem is to find a chirally invariant
description of the right side in (\ref{currentalgebra}). What we propose
here is to use an effective Lagrangian description which describes
$\langle\alpha|O|\beta\rangle$ and
\emph{nearby} processes in a chiral invariant way. This Lagrangian could
have also imaginary coefficients if that is needed to describe the
nearby underlying processes.

\begin{figure}
\setlength{\unitlength}{1pt}
\begin{center}
\setlength{\unitlength}{0.8pt}
\begin{picture}(100,100)
\SetScale{0.8}
\SetWidth{3}
\Line(0,100)(20,80)
\Line(20,80)(20,20)
\Line(0,0)(20,20)
\Line(20,80)(80,80)
\Line(20,20)(80,20)
\Line(20,20)(80,80)
\Line(80,80)(100,100)
\Line(80,20)(100,0)
\SetWidth{1.}
\Curve{(20.,80.)(50.,95.)(80.,80.)}
\Line(80,80)(80,20)
\end{picture}
\raisebox{50\unitlength}{$\Rightarrow$}
\setlength{\unitlength}{0.8pt}
\begin{picture}(100,100)
\SetScale{0.8}
\SetWidth{3}
\Line(0,100)(20,80)
\Line(20,80)(20,20)
\Line(0,0)(20,20)
\Line(20,80)(80,80)
\Line(20,20)(80,20)
\Line(20,20)(80,80)
\Line(80,80)(100,100)
\Line(80,20)(100,0)
\SetWidth{1.}
\Line(20.,80.)(40,100)
\Line(60,100)(80,80)
\Line(80,80)(100,60)
\Line(100,40)(80,20)
\end{picture}
\raisebox{50\unitlength}{$\Rightarrow$}
\setlength{\unitlength}{0.8pt}
\begin{picture}(100,100)
\SetScale{0.8}
\SetWidth{3}
\Line(0,70)(50,50)
\Line(0,30)(50,50)
\Line(50,50)(100,70)
\Line(50,50)(100,30)
\SetWidth{1.}
\Line(50,50)(30,100)
\Line(50,50)(70,100)
\Line(50,50)(30,0)
\Line(50,50)(70,0)
\Vertex(50,50){8}
\end{picture}
\raisebox{50\unitlength}{$\Rightarrow$}
\setlength{\unitlength}{0.8pt}
\begin{picture}(100,100)
\SetScale{0.8}
\SetWidth{3}
\Line(0,70)(50,50)
\Line(0,30)(50,50)
\Line(50,50)(100,70)
\Line(50,50)(100,30)
\SetWidth{1.}
\Oval(50,75)(25,15)(0)
\Oval(50,25)(25,15)(0)
\Vertex(50,50){8}
\end{picture}
\end{center}
\caption{An example of the argument used. The thick lines contain a
large momentum, the thin lines a soft momentum. 
Left: a general Feynman diagram with hard and soft lines.
Middle-left: we cut the soft lines to remove the soft singularity.
Middle-right: The contracted version where the hard part is
assumed to be correctly described
by a ``vertex'' of an effective Lagrangian.
Right:
the contracted version as a loop diagram. This is expected to reproduce
the chiral logarithm of the left diagram.}
\label{fig:power}
\end{figure}
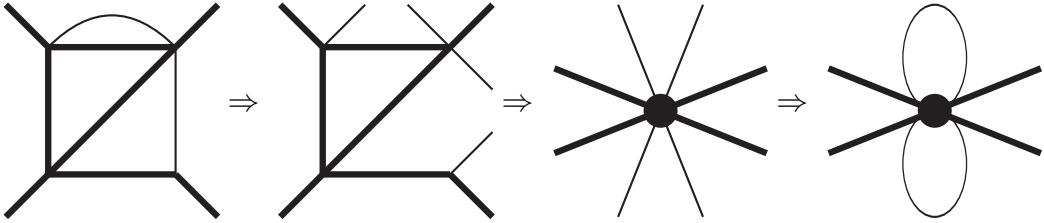
\noindent{\bf 2.)} For a general loop calculation, we expect that the hard part,
can be described
by an effective Lagrangian as long as none of the external momenta changes
very much. We take a Feynman diagram at a particular configuration of the
internal and external hard momenta. We cut the soft lines which are
repsonsible for the chiral logarithms and possibly other soft singularities.
The remainder is analytic in the soft quantities and should be describable
by an effective Lagrangian.
This is illustrated in Fig.~\ref{fig:power} and is
essentially the analysis of possible infrared divergences
as discussed in Sect.~8.3.1 in \cite{IZ}.
Related thoughts can be found in \cite{BGT} and in the work
on asymptotic expansions of loop integrals \cite{BS} and
in the first study of baryon ChPT \cite{GSS}. This effective Lagrangian
should then provide a sufficiently complete description of the process in
the neighbourhood of $\langle\alpha|O|\beta\rangle$, including extra
soft pions. Finding a complete
description in the relevant neighbourhood is thus the crux.
For the case of $K_{\ell3}$ decays at a general $q^2$ this was accomplished
in \cite{Flynn}
by showing that matrix-elements of higher order operators were related
to the matrix-elements of the lowest order operator to the order needed.

\noindent{\bf 3.)} Let us generalize the argument of \cite{Flynn} to the case at hand, 
$K\to2\pi$
decays. We look at matrix-elements of the type
$\langle\pi(p_1)\pi(p_2)|O|K(p_K)\rangle$ where $O$ is any of the operators
in $\lag_{1/2}$ or with a higher number of derivatives. We show here
that these matrix-elements are all proportional to the lowest order one
up to terms of order $\hat m$ times order one coefficients.

We will formulate the discussion in terms of the expansion in powers of $M^2$,
the lowest order pion mass.
The lowest order for $K\to\pi\pi$ in this counting
is order 1, then $M^2$ (plus logarithms), $M^4$,\ldots.
The combinations of hard momenta are $p_1^2 = p_2^2 = M_\pi^2$,
$p_K^2 = M_K^2$ and $p_1.p_K = p_2.p_K = M_K^2/2$.
Neither of the masses has a chiral logarithm of the type
$\logm=M^2\log(M^2)$.

Terms which contain powers of $\chi$ will not contribute to the order
1 or $\logm$ but only start at $M^2$. We thus need to look only
at terms with derivatives $\nabla_\mu$ or $u_\mu$.
Lorentz indices always come in pairs.

 {\bf (a)} Let us first look at the case where
both derivatives in the pair are from $\nabla_\mu$.
If the derivative hits a soft pion,
the underlying soft part of the loop integrals is 
$\int d^d p\, p_\mu/(p^2-M^2)$ which contributes no terms of order
$\logm$. So the only parts that can contribute are when the extra
derivatives both hit either the kaon or the two hard pions,
we will in the below thus always only consider the hard particles.
All options of how a pair of derivatives hit the
hard particles can be related to the lowest order term up to terms
of order $\logm$.

First, if both derivatives hit the same hard particle, it produces their
mass which contains no extra \logm as mentioned above.
Second, if they hit both pions, we can perform
a partial integration where only one derivative hits a pion and the other
the kaon plus mass term contributions. So we only need to consider
the case when one derivative hits a pion and the other the Kaon.
Third: $K\to\pi\pi$ is symmetric under the interchange of the pions,
so if we have a term with one derivative of the pair hitting the kaon and
the second derivative a pion,
there must thus be an identical term with the second derivative
hitting the other pion,
the pion momenta in this form are thus always $p_1+p_2$ but that means
that that derivative can always be moved by partial integration
to the kaon as well and turned into
a kaon mass. This takes care of all terms with extra powers of
$\nabla_\mu\ldots\nabla_\mu$. 

{\bf(b)} What happens now with terms with $u_\mu$,
where the derivatives must be on the hard pions. The remaining
terms are those of the type $E_2$, $E_3$ or $E_8$ in (\ref{lag1/2}).
These can all be related to the $E_1$ term up to order $M^2$.
We use the identity
\be
\partial_\mu\left(\tau_{1/2} \tilde K\right) =
 \frac{1}{2}\tau_{1/2} u_\mu \tilde K + \tau_{1/2} \nabla_\mu \tilde K\,,
\ee
valid for any $\tilde K$ transforming as
$\tilde K\longrightarrow h\tilde K$.
The matrix element of a total derivative vanishes since $p_1+p_2=p_K$.
Using $\tilde K= u^\mu K$ and $\tilde K = \nabla^\mu K$ we get
\ba
0 &=& \frac{1}{2}\tau_{1/2} u_\mu u^\mu K + \tau_{1/2}\nabla_\mu u^\mu K
+ \tau_{1/2} u_\mu \nabla^\mu K\,,
\nonumber\\
0 &=& \frac{1}{2}\tau_{1/2} u_\mu \nabla^\mu K + \tau_{1/2} \nabla_\mu\nabla^\mu K\,.
\ea
This shows that the $E_2$ and $E_3$ terms can be reduced to the $E_1$ term.
The $E_8$ term can also be removed, perform a partial integration
on one of the $\nabla_\mu$ hitting the Kaon. This produces either 
a $\nabla_\mu u^\mu$ which is of order $M^2$ or a $\nabla_\mu u_\nu$.
But in the latter case we can use that $\nabla_\mu u_\nu = \nabla_\nu u_\mu
+ f_{-\mu\nu}$ \cite{BCE} where the extra term vanishes for zero external
fields as is the case for $K\to\pi\pi$. The remainder is then of a form already discussed.
We have thus shown that for $K\to\pi\pi$ matrix elements all operators
have matrix elements that up to terms of order $M^2$ are proportional to
the lowest order operator.

{\bf (c)} The same type of arguments goes through for all
 $\Delta I=3/2$ operators.
We can also show that the terms with $D_1$ and $D_2$ in
(\ref{lag3/2}) are equivalent in the same way by considering
$\partial_\mu\left(\tau^{ij}_k (u^\mu)_i^{\phantom{i}k} K_j\right)$.

\noindent{\bf 4.)} The above argument does not work for relating $K\to2\pi$
to $K\to3\pi$ in general.
However the principle can again be applied if one of the pions
in $K\to3\pi$ is soft and the other two hard and in a momentum
configuration similar to $K\to2\pi$.
We have not checked whether additional operators
can already occur at lowest order for this case.

\noindent{\bf 5.)} The type of arguments presented above are
clearly applicable to many more
processes with hard momenta, in particular we expect that they can be applied
to matrix-elements needed for $B$ and $D$ decays as well, but again, we have
not performed such an analysis.

\section{The one-loop calculation for $K\to\pi\pi$}
\label{calculation}

There are three measured decays $K\to\pi\pi$: $ K_S \rightarrow \pi^0 \pi^0$,
 $K_S \rightarrow \pi^+ \pi^-$ and  $K^+ \rightarrow \pi^0 \pi^+ $  and
their charge conjugates. 
$K_S=\frac{1}{\sqrt{2}}(K^0 - \bar K^0)$ is the even CP eigenstate
and $K_L=\frac{1}{\sqrt{2}}(K^0 + \bar K^0) $ is an odd eigenstate.
The amplitudes for the three decays can be
written in terms of the $\Delta I=1/2$ and $3/2$ amplitudes $A_0$ and $A_2$.
\ba
A[K_S \rightarrow \pi^0 \pi^0] &=& \sqrt{\frac{2}{3}}A_0 -\frac{2}{\sqrt{3}}A_2
\,,
\nonumber \\
A[K_S \rightarrow \pi^+ \pi^-] &=&\sqrt{\frac{2}{3}}A_0 +\frac{2}{\sqrt{3}}A_2
\,,
 \nonumber \\
 A[K^+ \rightarrow \pi^0 \pi^+ ] &=& \frac{\sqrt{3}}{2} A_2\,.
\ea

\begin{figure}
\begin{center}
\setlength{\unitlength}{0.5pt}
\begin{picture}(100,120)(0,-40)
\SetScale{.5}
\SetWidth{2}
\Line(0,50)(50,50)
\Line(50,50)(100,0)
\Line(50,50)(100,100)
\GBoxc(50,50)(10,10){0}
\Text(50,-40)[b]{(a)}
\end{picture}
\hskip1cm
\setlength{\unitlength}{0.5pt}
\begin{picture}(100,120)(0,-40)
\SetScale{.5}
\SetWidth{2}
\Line(0,50)(50,50)
\Line(50,50)(100,0)
\Line(50,50)(100,100)
\Line(50,50)(50,0)
\Vertex(50,50){5}
\GBoxc(50,0)(10,10){0}
\Text(50,-40)[b]{(b)}
\end{picture}
\end{center}
\caption{\label{figtree}
Diagrams contributing to $K\rightarrow \pi \pi$ at tree level. 
A black box indicates a vertex from the weak Lagrangian,
(\ref{lag1/2}) or (\ref{lag3/2}),
and a black circle represent a vertex from the strong Lagrangian,
 (\ref{pilagrangian}) or (\ref{Klagrangian}). }
\end{figure}
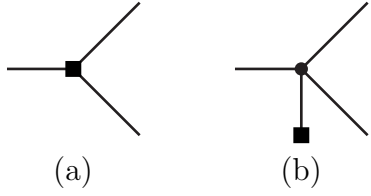
The tree level diagrams are shown in Fig.~\ref{figtree} and lead to
\ba
\label{resultLO}
A_0^{LO} &=& \frac{\sqrt{3}i}{2 F^2}\left[
 -\frac{1}{2} E_1+\left(E_2-4  E_3\right) \overline M_K^2+2 E_8 \overline M_K^4
+A_1 E_1\right]+\mathcal{O}(\logm)\,,
\nonumber\\
A_2^{LO} &=&
\sqrt{\frac{3}{2}}\frac{i}{F^2}
\left[\left(-2 D_1+D_2\right)\overline M_K^2\right]
+\mathcal{O}(\logm)\,.
\ea
We have kept here redundant terms to check explicitly the arguments of
Sect.~\ref{powercounting} and have dropped all terms of order $M^2$.
These come with new free coefficients as can be seen from the extra terms
in (\ref{lag1/2}) and (\ref{lag3/2}). The term with $A_1 E_1$ is the only
part coming from the tadpole diagram of Fig.~\ref{figtree}(b).

\begin{figure}
\begin{center}
\setlength{\unitlength}{0.5pt}
\begin{picture}(110,130)(0,-30)
\SetScale{.5}
\SetWidth{2}
\Line(0,50)(30,50)
\Line(70,50)(110,100)
\Line(70,50)(110,0)
\BCirc(50,50){20}
\GBoxc(30,50)(8,8){0}
\Vertex(70,50){5}
\Text(55,-30)[b]{(a)}
\end{picture}
\hskip7mm
\setlength{\unitlength}{0.5pt}
\begin{picture}(100,130)(0,-30)
\SetScale{.5}
\SetWidth{2}
\Line(50,50)(100,50)
\Line(0,50)(50,50)
\Line(50,50)(100,0)
\BCirc(50,70){20}
\GBoxc(50,50)(8,8){0}
\Text(50,-30)[b]{(b)}
\end{picture}
\hskip7mm
\setlength{\unitlength}{0.5pt}
\begin{picture}(110,140)(0,-40)
\SetScale{.5}
\SetWidth{2}
\Line(0,50)(30,50)
\Line(30,0)(30,50)
\Line(70,50)(110,100)
\Line(70,50)(110,0)
\BCirc(50,50){20}
\Vertex(70,50){5}
\Vertex(30,50){5}
\GBoxc(30,0)(8,8){0}
\Text(55,-40)[b]{(c)}
\end{picture}
\hskip7mm
\setlength{\unitlength}{0.5pt}
\begin{picture}(100,160)(0,-90)
\SetScale{.5}
\SetWidth{2}
\Line(0,50)(50,50)
\Line(50,50)(80,30)
\Line(50,50)(80,70)
\Line(50,50)(50,0)
\Line(50,0)(50,-40)
\BCirc(70,0){20}
\Vertex(50,50){4}
\Vertex(50,0){5}
\GBoxc(50,-40)(8,8){0}
\Text(50,-90)[b]{(d)}
\end{picture}
\hskip7mm
\setlength{\unitlength}{0.5pt}
\begin{picture}(80,160)(0,-70)
\SetScale{.5}
\SetWidth{2}
\Line(0,50)(50,50)
\Line(50,50)(80,30)
\Line(50,50)(80,70)
\Line(50,50)(50,0)
\BCirc(70,0){20}
\Vertex(50,50){5}
\Vertex(50,0){5}
\GBoxc(50,0)(8,8){0}
\Text(40,-70)[b]{(e)}
\end{picture}
\hskip7mm
\setlength{\unitlength}{0.5pt}
\begin{picture}(100,160)(0,-60)
\SetScale{.5}
\SetWidth{2}
\Line(50,50)(100,50)
\Line(0,50)(50,50)
\Line(50,50)(100,0)
\Line(50,50)(50,0)
\BCirc(50,70){20}
\Vertex(50,50){5}
\GBoxc(50,0)(8,8){0}
\Text(50,-60)[b]{(f)}
\end{picture}
\end{center}
\caption{\label{figloop}
Diagrams contributing to $K \rightarrow \pi \pi $  at one loop. 
Vertices as in Fig.~\ref{figtree}.}
\end{figure}
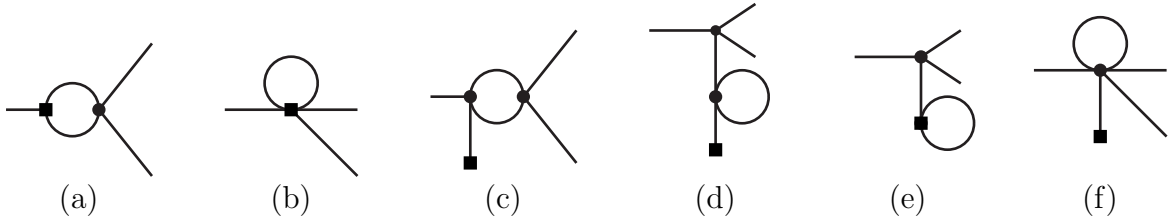
The one-loop diagrams are shown in Fig.~\ref{figloop} and there are in addition
contributions from wave-function renormalization. These diagrams are
not shown in Fig.~\ref{figloop}. Kaon wave-function renormalization has no
terms of order $\logm$ but pion wave-function renormalization contributes
to this order.

The tadpole diagrams
(c-f) do not contribute to $A_2$, only to $A_0$, as expected.
Diagrams (a) and (c) have $\pi\pi$ and $K\pi$ intermediate states.
All diagrams are nonzero but only a few have terms of order $\logm$.
Diagram (d) has no contribution but neither has (c). For diagram (a)
only the $\pi\pi$ intermediate state provides a contribution of order \logm.
The $K\pi$ intermediate state did contribute for $K_{\ell3}$.
The contributions from the different diagrams are given in Tab.~\ref{tabloop}.
\begin{table}
\begin{center}
\begin{tabular}{|c|c|c|}
\hline
Diagram & $A_0$ & $ A_2$ \\
\hline
\rule{0cm}{20pt} $Z$
&$-\frac{2F^2}{3}A_0^{LO}$ & $-\frac{2F^2}{3}A_2^{LO}$ \\[4pt]
(a) & $\sqrt{3}i\left(-\frac{1}{3}E_1+\frac{2}{3} E_2 \overline M_K^2\right)$&
 $\sqrt{\frac{3}{2}}i\left(-\frac{2}{3}D_2\overline M_K^2\right)$\\[4pt]
(b) & $\sqrt{3}i\left(-\frac{5}{96}E_1
 -\left(\frac{7}{48}E_2+\frac{25}{12} E_3\right) \overline M_K^2
  +\frac{25}{24}E_8 \overline M_K^4\right)$ &
 $\sqrt{\frac{3}{2}}i
  \left(-\frac{61}{12}D_1+\frac{77}{24}D_2\right)\overline M_K^2$
\\[4pt]
(e) & $\sqrt{3}i\frac{3}{16} A_1 E_1$ & \\[4pt]
(f) & $\sqrt3i\left(\frac{1}{8}E_1+\frac{1}{3}A_1 E_1\right)$ & \\[4pt]
\hline
\end{tabular}
\end{center}
\caption{\label{tabloop} The coefficients of $\overline A(M^2)/F^4$ 
in the contributions to
$A_0$ and $A_2$ from the different diagrams in Fig.~\ref{figloop}.
$Z$ denotes the part from wave-function renormalization.}
\end{table}
Putting all the diagrams together, we do indeed find a universal
coefficient for all the \logm\ terms:
\ba
\label{resultNLO}
A_0^{NLO} &=& A_0^{LO}\left(1+\frac{3}{8F^2}\overline A(M^2)\right)+
\lambda_0 M^2+\mathcal{O}(M^4)\,,
\nonumber\\
A_2^{NLO} &=& A_2^{LO}\left(1+\frac{15}{8F^2}\overline A(M^2)\right)+
\lambda_2 M^2+\mathcal{O}(M^4)\,.
\ea
Since we included redundant terms this also provides a check of the
arguments given in Sect.~\ref{powercounting}.

For a reasonable choice of $M^2$ and $\mu^2$ $\overline{A}(M^2)$
is positive, the result (\ref{resultNLO}) goes in the opposite direction
required for the $\Delta I=1/2$ rule, however if lattice calculations of
$K\to\pi\pi$ directly at sufficiently low $M^2$ and physical $m_s$
become available (\ref{resultNLO}) can be used to perform the extrapolation
to the physical pion mass.
 
Our result (\ref{resultNLO}) is
not directly related to the final state interaction of the two pions (FSI),
the main effect from that is dependent on $s_\pi$ $(=m_K^2)$, 
not on the pion mass
and would survive in the limit $M^2\to0$ keeping $m_K^2$ finite.
FSI effects in $K\to\pi\pi$ have been analyzed by many authors,
see \cite{FSI} and references therein.
It should be kept in mind as well that we have not used any soft pion
approximation for the two pions present in the decay $K\to\pi\pi$,
only for any additional pions relevant for the nonanalytic behaviour in $M^2$.

\section{Comparison with the three-flavour result}
\label{threeflavour}

Three flavour ChPT has been used a lot for $K\to\pi\pi$ decays.
The isospin conserving calculations were done first in \cite{KMW1} and 
recalculated in \cite{BPP98} and \cite{BDP}. The calculations of the
logarithmic terms go back even further.
By taking the published expressions from \cite{BPP98} and performing the
limit $M^2\to 0$ carefully we can compare with our results of
two-flavour ChPT. The lowest order result there reads
\ba
\label{result3LO}
A_0^{(3)LO} &=& -\frac{i\sqrt{6}CF_0^4}{\overline F_K F^2}
\left(G_8+\frac{1}{9}G_{27}\right)\overline M_K^2\,,
\nonumber\\
A_2^{(3)LO} &=& -\frac{i10\sqrt{3}CF_0^4}{9\overline F_K F^2}
G_{27}\overline M_K^2\,,
\ea
and can be used to determine the two-flavour LECs in terms of
the three-flavour LECs by comparing (\ref{resultLO}) and (\ref{result3LO}).

We can now check whether the full three-flavour one-loop result also
produces the same \logm\ terms as were calculated here.
To do this one must take into account that the lowest order
result in \cite{BPP98} was expressed in terms of $F_K$ and $F_\pi$.
To compare with (\ref{resultNLO}) we thus need to take into account
the \logm\ terms present in (\ref{Fpi}) and (\ref{FK}).
Doing this we do obtain the same result as in (\ref{resultNLO}) with
$A_i^{LO}$ replaced by $A^{(3)LO}_i$.
Note that the corrections terms $\lambda_i M^2$ in three-flavour perturbation
are also free at NLO there since they contain undetermined LECs.

\section{Conclusions}
\label{conclusions}

We have argued that it is possible to have a ``hard pion'' ChPT
and provided explicit arguments that in nonleptonic $K\to2\pi$ the
correction of order \logm\ is calculable. The arguments given in
Sect.~\ref{powercounting} provide the main basis of this work.
We then performed the calculation explicitly in Sect.~\ref{calculation}
keeping some of the redundant terms and showed that the arguments also
worked out in the explicit calculation. Equation (\ref{resultNLO})
is the main analytical result of this paper and should be useful
for extrapolating direct lattice calculations of $K\to\pi\pi$
to the physical pion mass.
As a final check we performed the matching to the known three-flavour
one-loop ChPT result.

\section*{Acknowledgements}

This work is supported in part by the European Commission RTN network,
Contract MRTN-CT-2006-035482  (FLAVIAnet), 
European Community-Research Infrastructure
Integrating Activity
``Study of Strongly Interacting Matter'' (HadronPhysics2, Grant Agreement
n. 227431)
and the Swedish Research Council.
FORM \cite{FORM} was used for the calculations.


\begin{thebibliography}{99}


\bibitem{Weinberg0}
  S.~Weinberg,
  Physica A {\bf 96} (1979) 327.

\bibitem{GL0}
  J.~Gasser and H.~Leutwyler,
  Nucl.\ Phys.\  B {\bf 250} (1985) 465.

\bibitem{GL1}
  J.~Gasser and H.~Leutwyler,
  Annals Phys.\  {\bf 158} (1984) 142.

\bibitem{KMW1}
  J.~Kambor, J.~H.~Missimer and D.~Wyler,
  Phys.\ Lett.\  B {\bf 261} (1991) 496.

\bibitem{BPP98}
  J.~Bijnens, E.~Pallante and J.~Prades,
  Nucl.\ Phys.\  B {\bf 521} (1998) 305
  [arXiv:hep-ph/9801326].

\bibitem{BDP}
  J.~Bijnens, P.~Dhonte and F.~Persson,
  Nucl.\ Phys.\  B {\bf 648} (2003) 317
  [arXiv:hep-ph/0205341].

\bibitem{LC}
  S.~Li and N.~H.~Christ,
  arXiv:0812.1368 [hep-lat].

\bibitem{Roessl}
  A.~Roessl,
  Nucl.\ Phys.\  B {\bf 555} (1999) 507
  [arXiv:hep-ph/9904230].

\bibitem{Ouellette}
  S.~M.~Ouellette,
  arXiv:hep-ph/0101055.

\bibitem{Frink}
  M.~Frink, B.~Kubis and U.~G.~Meissner,
  Eur.\ Phys.\ J.\  C {\bf 25} (2002) 259
  [arXiv:hep-ph/0203193].

\bibitem{Allton}
  C.~Allton {\it et al.}  [RBC-UKQCD Collaboration],
  Phys.\ Rev.\  D {\bf 78} (2008) 114509
  [arXiv:0804.0473 [hep-lat]].

\bibitem{Flynn}
  J.~M.~Flynn and C.~T.~Sachrajda  [RBC Collaboration and UKQCD
                 Collaboration],
  Nucl.\ Phys.\  B {\bf 812} (2009) 64
  [arXiv:0809.1229 [hep-ph]].

\bibitem{Celis}
  A. Celis, LU TP 09-12, Master thesis Lund University.

\bibitem{Lellouch}
 L.~Lellouch,
  arXiv:0902.4545 [hep-lat].

\bibitem{BCE}
  J.~Bijnens, G.~Colangelo and G.~Ecker,
  JHEP {\bf 9902} (1999) 020
  [arXiv:hep-ph/9902437].

\bibitem{IZ} C. Itzykson and J.-B. Zuber, Quantum Field Theory,
McGraw-Hill, New York, 1980.

\bibitem{BGT}
  J.~Bijnens, P.~Gosdzinsky and P.~Talavera,
  JHEP {\bf 9801} (1998) 014
  [arXiv:hep-ph/9708232].

\bibitem{BS}
  M.~Beneke and V.~A.~Smirnov,
  Nucl.\ Phys.\  B {\bf 522} (1998) 321
  [arXiv:hep-ph/9711391].

\bibitem{GSS}
  J.~Gasser, M.~E.~Sainio and A.~Svarc,
  Nucl.\ Phys.\  B {\bf 307} (1988) 779.

\bibitem{FSI}
  T.~N.~Truong,
  Phys.\ Lett.\  B {\bf 207} (1988) 495;
  W.~A.~Bardeen, A.~J.~Buras and J.~M.~Gerard,
  Phys.\ Lett.\  B {\bf 192} (1987) 138;
  V.~Antonelli, S.~Bertolini, M.~Fabbrichesi and E.~I.~Lashin,
  Nucl.\ Phys.\  B {\bf 469}, 181 (1996)
  [arXiv:hep-ph/9511341];
  J.~Bijnens and J.~Prades,
  JHEP {\bf 0006} (2000) 035
  [arXiv:hep-ph/0005189];
  E.~Pallante and A.~Pich,
  Phys.\ Rev.\ Lett.\  {\bf 84} (2000) 2568
  [arXiv:hep-ph/9911233],
  Nucl.\ Phys.\  B {\bf 592} (2001) 294
  [arXiv:hep-ph/0007208];
  M.~Buchler, G.~Colangelo, J.~Kambor and F.~Orellana,
  Phys.\ Lett.\  B {\bf 521} (2001) 22
  [arXiv:hep-ph/0102287].

\bibitem{FORM}
  J.~A.~Vermaseren,
  math-ph/0010025.


\end{thebibliography}
\end{document}